\documentclass[aps,prb,twocolumn,showpacs,superscriptcite]{revtex4-1}
\usepackage{graphics}
\usepackage{graphicx}
\usepackage{dcolumn}
\usepackage{bm}
\usepackage{float}
\usepackage{subfigure}
\usepackage{todonotes}
\usepackage{amsfonts}
\usepackage{amssymb}
\begin{document}

\title{Physical properties of Ce$_{3-x}$Te$_4$ below room temperature}

\author{Andrew F. May}
\affiliation{Materials Science and Technology Division, Oak Ridge National Laboratory, Oak Ridge, TN 37831}
\author{Michael A. McGuire}
\affiliation{Materials Science and Technology Division, Oak Ridge National Laboratory, Oak Ridge, TN 37831}
\author{Claudia Cantoni}
\affiliation{Materials Science and Technology Division, Oak Ridge National Laboratory, Oak Ridge, TN 37831}
\author{Brian C. Sales}
\affiliation{Materials Science and Technology Division, Oak Ridge National Laboratory, Oak Ridge, TN 37831}

\date{\today}

\begin{abstract}
The physical properties of polycrystalline Ce$_{3-x}$Te$_4$ were investigated by measurements of the thermoelectric properties, Hall coefficient, heat capacity, and magnetization.  The fully-filled, metallic $x$=0 compound displays a soft ferromagnetic transition near 4\,K, and analysis of the corresponding heat capacity anomaly suggests a doublet ground state for Ce$^{3+}$.  The transition is suppressed to below 2\,K in the insulating $x$=0.33 composition, revealing that magnetic order in Ce$_{3-x}$Te$_4$ is driven by an RKKY-type interaction. The thermoelectric properties trend with composition as expected from simple electron counting, and the transport properties in Ce$_3$Te$_4$ are observed to be similar to those in La$_3$Te$_4$.  Trends in the low temperature thermal conductivity data reveal that the phonons are efficiently scattered by electrons, while all compositions examined have a lattice thermal conductivity near 1.2\,W/m/K at 200\,K.
\end{abstract}


\maketitle

\section{Introduction}

Rare-earth ($R$) chalcogenides ($Ch$) often adopt the cubic Th$_3$P$_4$ structure-type,\cite{WoodReview} and these compounds support a compositional range ($R_{3-x}$$Ch_4$) associated with $R$ vacancies ($0 \le x \le 1/3$).  These materials are promising for high temperature thermoelectric application due to their desirable transport properties, as well as thermal stability. The free electron concentration $n$ is directly proportional to the rare-earth content, with the number of free electrons per formula unit given by $1-3x$.  The transport properties vary from metallic to insulating as $x$ increases from 0 to $\frac{1}{3}$, and thus the inherent $n$-type thermoelectric performance can be rationally optimized via composition without introducing additional elements.  Early works on these materials identified promising performance above 1000\,K, and were nicely summarized by Wood.\cite{WoodReview}  

Recent work on La$_{3-x}$Te$_4$ generated a renewed interest in these materials,\cite{MayLaTe} in part due to the possibility of using Yb$_{14}$MnSb$_{11}$ as the $p$-type leg in a high-temperature thermoelectric couple.\cite{Brown14111,ToberAl}  It has been shown that while a relatively simple optimization scheme achieves large thermoelectric efficiency in La$_{3-x}$Te$_4$, a single band model is not an accurate description of the conduction band for the optimum compositions.  Rather, the conduction band is composed of lighter bands at the band edge and heavier bands roughly 0.3\,eV higher in energy.\cite{BandsLa3Te4} In Ce$_{3-x}$Te$_4$, the heavy bands are calculated to be $\sim$0.15\,eV from the band edge.\cite{Ce3Te4APL}  This results in a relatively broad range of compositions that yield large thermoelectric performance, and is thus a desirable band structure from an application standpoint.  Thus, it is not surprising to find a wide range of compositions suggested as the optimum thermoelectric composition in the literature.\cite{WoodReview,Re3S4Gadzhiev} Chemical substitutions have been considering for facilitating the optimization process\cite{LaYbS88,LYT} or examining pertinent electron and phonon scattering mechanisms.\cite{LTS}

To a large extent, the recent interest in La$_{3-x}$Te$_4$ is associated with the demonstration that it can be reproducibly synthesized in large quantities.  This is achieved using the low-temperature, solid-state synthesis route of ball milling (mechanical alloying).  High-melting temperatures and complications in the phase diagrams motivate the desire for a low-temperature synthesis route in these materials.  This process begins with the pure elements, and in the case of La$_{3-x}$Te$_4$ and Ce$_{3-x}$Te$_4$ produces single phase powders that are then hot pressed to yield dense samples.  It is important to stress that in these materials ball milling is utilized for synthesis, and not for particle size reduction.  While the powders produced contain small grains, the materials themselves have low thermal conductivities and low mobilities,\cite{Zhuze70,WoodReview} and thus the grain size generally has little influence on the transport properties when the grain boundaries are not oxidized.  This is particularly true at the high temperatures where these materials would operate in a thermoelectric device.

Wang et al. suggested that a large peak in the density of states above the Fermi level may lead to anomalously large thermoelectric efficiency for the metallic Ce$_3$Te$_4$ ($x=0$).\cite{Ce3Te4APL}  At the materials level, thermoelectric efficiency is quantified via $zT = \frac{\alpha^2T}{\rho\kappa}$, where $\alpha$ is the Seebeck coefficient, $\rho$ the electrical resistivity, and $\kappa$ the thermal conductivity.  Most state-of-the art thermoelectric materials have $zT$ ranging from 1 to 1.5 at the hot-side operating temperature.  La$_{3-x}$Te$_4$ reaches $zT$=1.2 at 1273\,K for $x\sim$0.2, and it was proposed that $zT>$10 may be achievable in Ce$_3$Te$_4$.  We note that high temperature transport properties of Ce$_{3-x}$Te$_4$ were investigated by Zhuze et al. in 1970,\cite{Zhuze70} but the temperature-dependent data for Ce$_{3-x}$Te$_4$ were not shown and their report stated that all $R_{3-x}$Te$_4$ examined behaved qualitatively similar.  

One primary goal for this study is to examine the thermoelectric efficiency of Ce$_{3-x}$Te$_4$ and determine whether or not a large thermoelectric efficiency is likely.  To that end, polycrystalline samples have been prepared and their thermoelectric properties were measured below room temperature.  The results are compared to those for La$_3$Te$_4$, which has no tendency toward mixed valency or Kondo interactions, and does not have any $f$ electrons that might influence transport.  While these materials would be utilized at high temperatures ($T>1000\,K$), the transport properties are sensitive to band structure features at low temperatures and thus this basic characterization is expected to provide significant insight into the general thermoelectric behavior of Ce$_{3-x}$Te$_4$.  Specific heat and magnetization measurements have also been performed to complete the characterization of basic physical properties.  Taken together, these results indicate that transport in Ce$_{3-x}$Te$_4$ is not strongly influenced by the $f$-levels, and efficiency similar to that in La$_{3-x}$Te$_4$ will likely be observed at high temperatures.  It is also shown that the cerium moments order ferromagnetically near 4\,K for the $x=0$ compound, and this ordering is suppressed as cerium vacancies are introduced.  To probe the influence of disorder and charge carriers on the magnetism, a sample of Ce$_{2.67}$La$_{0.33}$Te$_4$ was prepared because this composition contains a similar degree of Ce-site disorder as Ce$_{2.67}$Te$_4$, but is metallic not insulating.  Taken together, the magnetization measurements demonstrate that the observed ferromagnetic ordering of Ce moments is clearly driven by Ruderman-Kittel-Kasuya-Yosida (RKKY) indirect exchange interactions.

\section{Methods}

Polycrystalline samples of Ce$_{3-x}$Te$_4$ with $x=0,0.15,0.25,0.33$, as well as Ce$_{2.67}$La$_{0.33}$Te$_4$, were prepared via ball milling and hot pressing.  Cerium and/or lanthanum ingots (Ames laboratory) were cut or filed and combined with tellurium pieces (Alfa, 99.999\%) inside a helium glove box. A SPEX SamplePrep 8000-series Mixer/Mill was utilized, and milling occurred with a tungsten carbide lined vial utilizing two tungsten carbide balls. Total masses of 5-7\,g were loaded, and six hours of milling were performed during approximately 8-9\,h.  Hot pressing occurred in a graphite furnace and a high density graphite die was utilized with a grafoil liner to isolate the sample. The graphite die had an inner diameter of approximately 10\,mm and an 8.8\,mm diameter graphite rod compressed the sample with a maximum force of 500\,kg.  Sintering generally occurred by $\sim$1223\,K, while rapidly ramping the temperature, as observed via the ram travel. The samples were annealed at 1373\,K for two hours, with the force being completely removed after one hour.  Samples were cooled to 973\,K at 100\,K/h, and then the furnace was turned off.  Geometric densities were greater than 95\% of the theoretical densities calculated assuming nominal compositions and the refined lattice parameters.  The cylindrical samples were cut using a low speed diamond saw with glycerol for lubrication, and bar shaped samples were prepared for transport property measurements.  The samples were found to oxidize slowly in air at ambient conditions, and were stored in a glove box.  Measurements in an ULVAC ZEM-3 failed due to oxidation below $\sim$700\,K, and thus it is clear the samples are highly air sensitive at elevated temperatures.  For comparison, a sample of La$_3$Te$_4$ from Ref. \citenum{OlivierLaTe} that was produced through similar methods was cut for thermal transport measurements after being stored under argon.

Sample quality was examined via powder x-ray diffraction using a PANalytical X'Pert Pro MPD with a Cu K$_{\alpha,1}$ monochromator.  Rietveld refinements were performed using the program FullProf.\cite{FullProf} Powder diffraction experiments performed after 12-24\,h of air exposure revealed a minor weakening of intensity, but no decomposition was observed.

\begin{figure}
	\centering
\includegraphics[width=3.in]{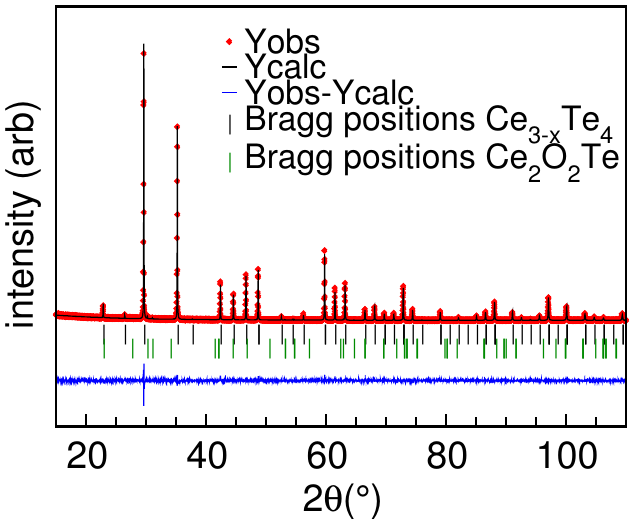}
\caption{(color online) Representative powder x-ray diffraction data are shown for a hot pressed sample of nominal composition Ce$_{2.85}$Te$_4$.  Only a very minor amount of the primary impurity phase (Ce$_2$O$_2$Te) is generally observed.}
	\label{fig:pxrd}
\end{figure}

\begin{table*}
  \caption{Selected parameters from refinements of powder x-ray diffraction data for Ce$_{3-x}$Te$_4$ and related compounds examined at room temperature. The Te position is at (x,x,x), and Ce(La) resides at (0.375,0,0.25). Refined fractional occupancies are shown for Ce(La), where the lower limit of $x=\frac{1}{3}$ would equal 0.889. The occupancy provided for the mixed Ce/La compound is the total site occupancy, which was obtained by refining Ce content while fixing the La occupation at its nominal 1/9.}
  \label{tab:pxrd}
  \begin{tabular}{lcccccc}
    \hline
   nominal composition         & Ce$_3$Te$_4$       & Ce$_{2.85}$Te$_4$   & Ce$_{2.75}$Te$_4$   &  Ca$_{2.67}$Te$_4$  &  Ce$_{2.67}$La$_{0.33}$Te$_4$ & La$_3$Te$_4$   \\
   \textit{a}(\AA)             & 9.53974(6)         & 9.53677(6)          &  9.53465(4)         & 9.53529(4)          &  9.54993(6)                   & 9.62603(5)     \\
   Ce(La) occupancy            & 0.96(1)            & 0.94(1)             &  0.90(1)            & 0.86(1)             &  0.97(1)                            & 0.97(1)        \\
   Te position                 & 0.07483(11)        & 0.07415(12)         &  0.07444(10)        & 0.07412(10)         &  0.07495(12)                  & 0.07518(9)     \\
   $\chi^2$               		 & 1.38               & 1.27                &  1.49               & 1.70                &  1.33                         & 1.34           \\
   $R_p$ / $R_{wp}$            & 8.36/10.9          & 9.12/12.1           &  8.89/11.5          & 9.19/12.2           &  9.13/12.2                    & 8.68/11.9      \\
   \hline
  \end{tabular}
\end{table*}

Particle size and crystallinity after hot pressing were investigated by transmission electron microscopy (TEM).  TEM characterization was carried out in a Philips CM200 microscope operating at 200\,kV and equipped with a Schottky field-emission gun. TEM specimens were obtained by sprinkling finely ground powders onto Cu grids coated with a holey carbon film.  Powders were ground in a He glove box, and were kept under He until the time of measurement to minimize exposure to air.

Transport measurements were completed in a Quantum Design Physical Property Measurement System (PPMS) and magnetization measurements were performed in a Quantum Design Magnetic Property Measurement System using a SQUID magnetometer.  Electrical contacts were made with DuPont 4929N silver paste.  During thermal transport measurements, silver epoxy (H20E Epo-Tek) was utilized to provide mechanical and thermal contact, and the silver paste provided improved electrical contacts.  For the semiconducting Ce$_{2.67}$Te$_4$ sample, gold was sputtered on the sample to further improve the contacts prior to electrical and thermal measurements.   Hall coefficients ($R_H$) were obtained on thinned samples (thickness less than 0.35\,mm) from a fit of the Hall resistance versus magnetic field, with maximum fields of $\pm$6\,T employed.  

\section{Results and Discussion}
\subsection{Chemical Characterization}

The as-milled materials produced powder x-ray diffraction (PXRD) scans consistent with the Th$_3$P$_4$ structure type, though peaks were very broad and of low intensity.  After hot-pressing, the PXRD scans on hand-ground powders revealed high-purity, well crystallized samples.  A typical diffraction scan and the corresponding refinement are shown in Figure \ref{fig:pxrd}.  Only very small amounts of the secondary phases can be identified in the powder diffraction data.  Ce$_2$O$_2$Te is the primary impurity phase, with a maximum refined weight fraction of 1.1\% (observed in the $x=0$ sample) and 1.9\% of La$_2$O$_2$Te was observed in the La$_3$Te$_4$, which was originally characterized in Ref \citenum{OlivierLaTe}.  In addition to this oxytelluride, CeTe (0.7 wt.\%) was observed in the sample with nominal composition Ce$_{2.67}$La$_{0.33}$Te$_4$ with a refined main phase purity of 98.9 wt.\%.  Refinements also revealed the expected trend in cerium concentration, as shown in Table \ref{tab:pxrd}.  Finally, the PXRD data for the La substituted sample reveal a lattice parameter that is nearly identical to that expected from Vegard's law assuming nominal composition and the end-member lattice parameters listed in Table \ref{tab:pxrd}.

\begin{table*}
\caption{Summary of properties in polycrystalline Ce$_{3-x}$Te$_4$.  Effective moments $\mu_{eff}$ were obtained assuming nominal compositions for Ce content, and fit results for both high temperature (HT = 30-300\,K) and low temperature (LT = 10-30\,K) are shown.  Curie temperatures $T_{\mathrm{C}}$ were defined as the mid-point of the steep rise in the induced moment upon cooling in an applied field of 50\,Oe.  Debye temperatures were obtained by fitting the heat capacity data between 20 and 200\,K to the Debye model.}\label{tab:props}
\begin{tabular}{ccccccccccc}
\hline
composition  & symbol &$n_H$  & $\rho$ & $\mu_H$ & $\alpha$ & $m^*_{SPB}$ & $\mu_{eff}$ (HT) & $\mu_{eff}$ (LT) & $T_{\mathrm{C}}$  & $\Theta_D$ \\
 -  & -   & 10$^{21}$cm$^{-3}$ &   m$\Omega$ cm &cm$^{2}$/V/s & $\mu$V/K & $m_e$ & $\mu_B$/Ce & $\mu_B$/Ce & K & K \\
\hline
Ce$_3$Te$_4$                    & $\circ$, $x=0$            & 4.6  & 0.34   &  4.0 & -22  & 3.0 & 2.52759         & 1.6922  & 3.7 & 159.8    \\
Ce$_{2.85}$Te$_4$               & $\diamond$, $x=0.15$      & 1.8  & 0.53   &  8.4 & -31  & 2.3 & 2.48874,2.58844 & 1.747   & 4.0 &          \\
Ce$_{2.75}$Te$_4$               & $\square$, $x=0.25$       & 0.84 & 0.91   &  8.2 & -35  & 1.5 & 2.55788         & 1.8128  & 2.6 & 168.17   \\
Ce$_{2.67}$Te$_4$               & $\triangledown$, $x=0.33$  &      & 3.5$\times$10$^4$ & &-530& & 2.55139        & 1.6678  &     & 172.633  \\
Ce$_{2.67}$La$_{0.33}$Te$_4$    & $\vartriangle$, La=0.33   & 4.94 & 0.30   &      & -21  & 3.0 & 2.67596         & 1.697   & 2.9 & 157.08   \\
La$_3$Te$_4$                    & $\times$                  &      & 0.36   &      & -21  &     & 2.67596         &         &     & 178.076  \\
\hline
\end{tabular}
\end{table*}

TEM was undertaken to examine particle size in the compositions Ce$_3$Te$_4$ and Ce$_{2.67}$Te$_4$.  After hand-grinding the hot-pressed samples in a He glove box, the samples were found to contain particles that were composed of many smaller, crystalline grains.  Both the average particle size and average crystallite size are found to be smaller in the $x=0.33$ sample than in the $x=0$ sample.  Average area(long-dimension) of the particles were observed to be 0.45\,$\mu$m$^2$(0.95\,$\mu$m) for $x=0$ and 0.17\,$\mu$m$^2$(0.65\,$\mu$m) for $x=0.33$.  For the powders viewed, the grain size in the $x=0$ sample was found to be $\sim$50\,nm in length as compared to a grain size of $\sim$10\,nm for the $x=0.33$ sample, though some particles were observed to contain grains as large as 500\,nm.  Grain sizes obtained by Scherrer analysis using X'Pert HighScore Plus were $\sim$105\,nm for $x=0$ and $\sim$120\,nm for $x=0.33$ (not considering instrument broadening or strain).

\begin{figure}
	\centering
\includegraphics[width=3.in]{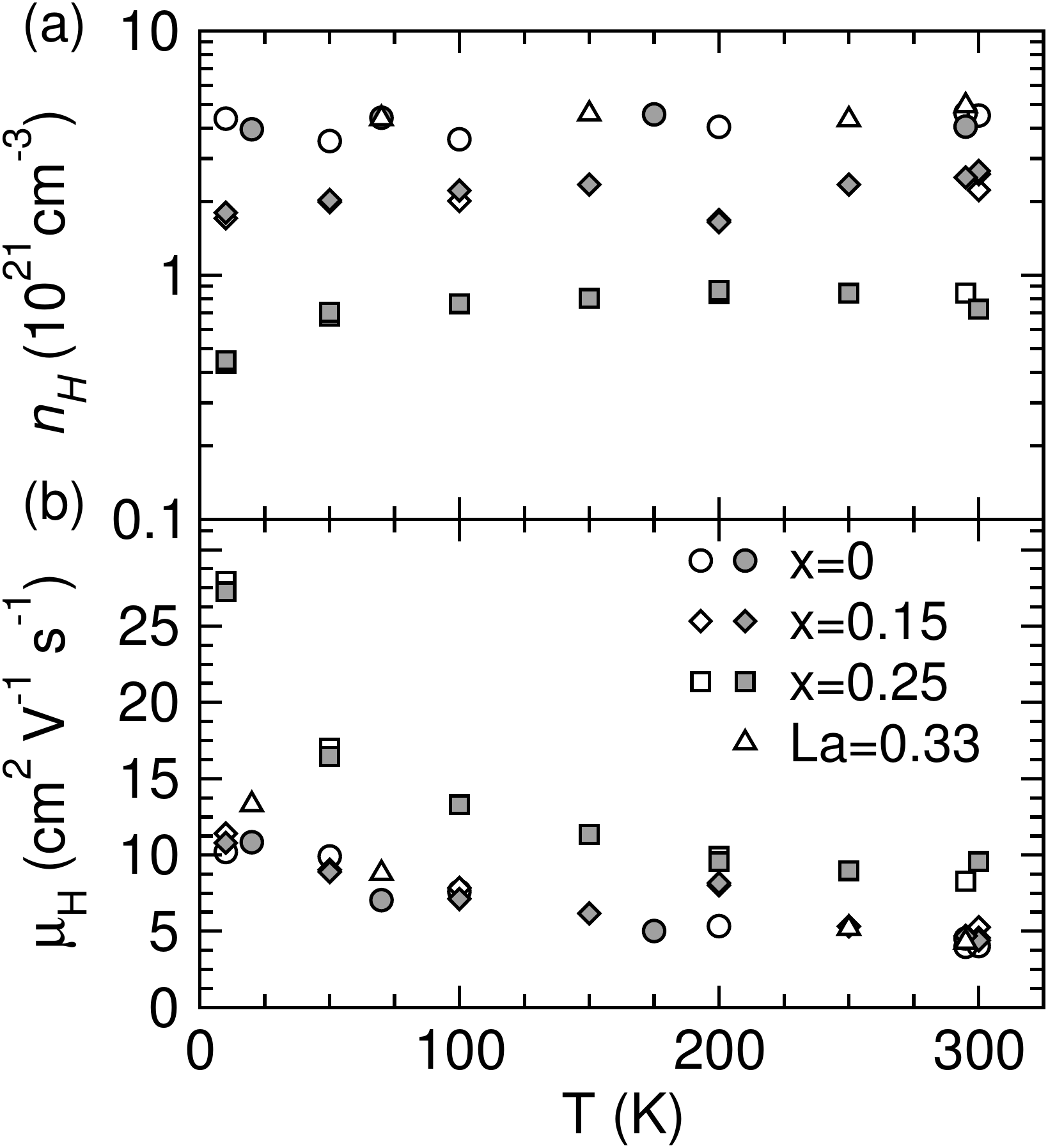}
\caption{Temperature dependence of the (a) absolute value of the Hall carrier density ($n$-type) and (b) the Hall mobility in metallic Ce$_{3-x}$Te$_4$ samples.  Data for two samples (different symbol filling) are shown for each composition to represent the distribution of compositions/properties across one hot-pressed specimen, and La=0.33 data are from the Ce$_{2.67}$La$_{0.33}$Te$_4$ sample.}
	\label{fig:hall}
\end{figure}

\subsection{Transport Properties}
 
The room temperature transport properties are summarized in Table \ref{tab:props}.  The trends observed are consistent with previous studies on $R_{3-x}$$Ch_4$ compounds, and the data for $x$=0 are consistent with the literature.\cite{Zhuze70} $R_{3-x}$$Ch_4$ compounds can be considered to be doped semiconductors, where the composition determines the `extrinsic' doping level with (1-3$x$) electrons per formula unit in the conduction band.  As a result, the carriers are not expected to freeze-out like in traditional doped semiconductors where carriers may or may not form impurity bands. This is confirmed via Hall coefficient ($R_H$) measurements, which revealed $n$-type conduction and relatively little temperature dependence as illustrated by the Hall carrier densities ($n_H$=1/$R_H e$) shown in Figure \ref{fig:hall}a.  As expected, the values of $n_H$ are relatively temperature independent and trend with cerium content.

\begin{figure}
	\centering
\includegraphics[width=3.in]{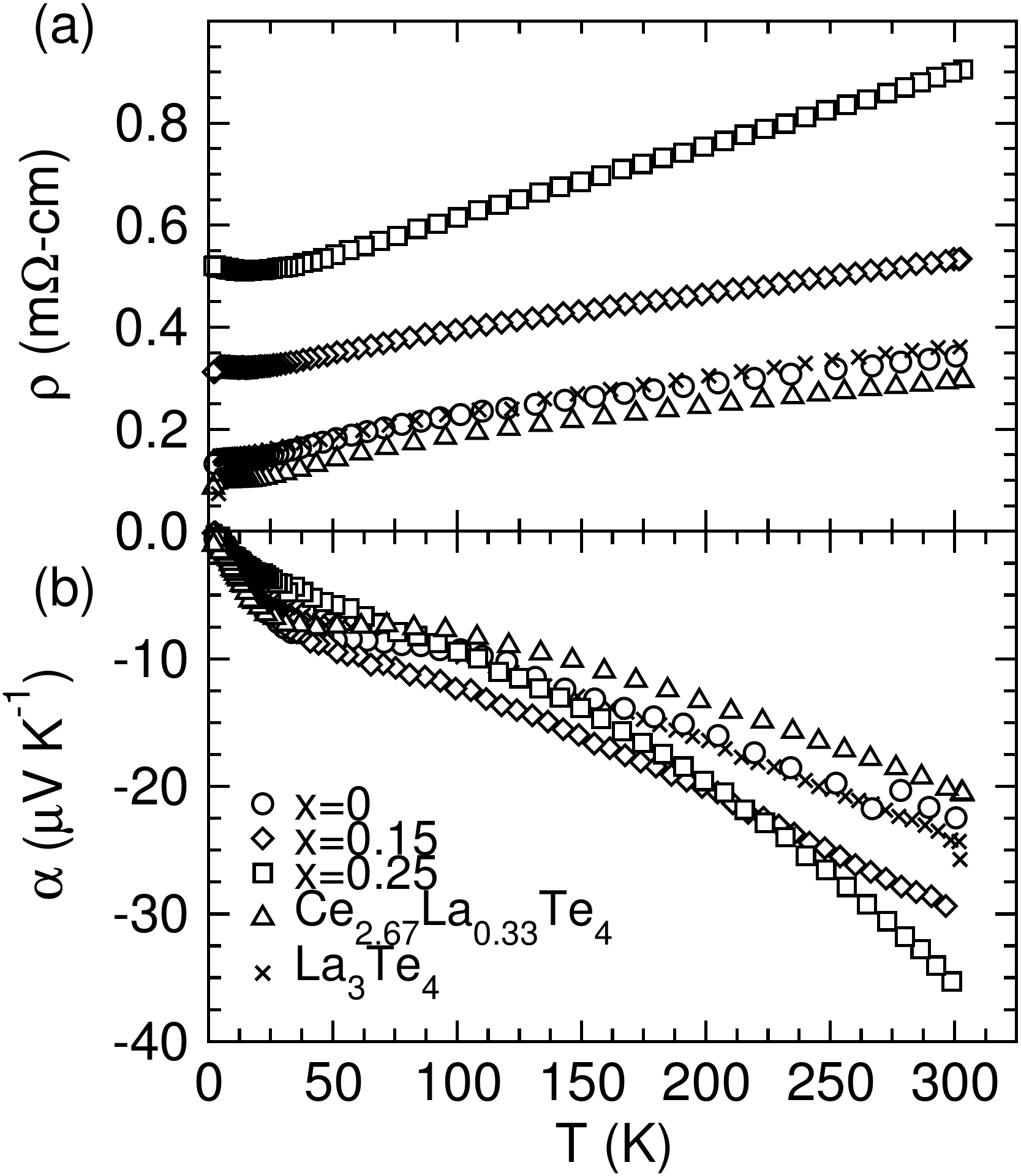}
\caption{(a) Electrical resistivity and (b) Seebeck coefficient of Ce$_{3-x}$Te$_4$ samples show metallic behavior for samples with $x<0.3$.  Data for Ce$_{2.67}$La$_{0.33}$Te$_4$ and La$_3$Te$_4$ are also included.}
	\label{fig:elec}
\end{figure}

The electrical resistivity $\rho$ of Ce$_{3-x}$Te$_4$ samples increases with increasing temperature for $x \le 0.25$, as shown in Figure \ref{fig:elec}a.  For the $x=0.25$ sample, a small upturn in $\rho$ is observed at low temperatures, possibly due to Anderson localization.\cite{CutlerMott}  Trends with composition are clear and as expected.  It has been pointed out that the electrical resistivity is perhaps the most `sensitive indicator of composition' in these $R_{3-x}$Te$_4$ materials (assuming similar syntheses).\cite{Ikeda82} 

As observed in Figure \ref{fig:elec}b, the Seebeck coefficients $\alpha$ are small in magnitude and negative in sign, consistent with the expected and observed carrier densities.  The magnitudes of $\alpha$ decrease with decreasing temperature, which is typical for itinerant electrons with a relatively fixed concentration.  The magnitudes also trend nicely with composition at room temperature.  Effective mass $m^*$ values have been calculated from the Seebeck and Hall coefficients, under the assumption of a simple parabolic band with a carrier relaxation time that is limited by acoustic phonon scattering.  The results are summarized in Table \ref{tab:props}, and the values for the $x=0$ composition are consistent with previous reports.  The decreased $m^*$ at lower $n_H$ is expected from the electronic structure,\cite{Ce3Te4APL} and was also observed in La$_{3-x}$Te$_4$.\cite{BandsLa3Te4}  For instance, consider the theoretical plot of $\alpha$ versus $n_H$ in La$_{3-x}$Te$_4$ (Fig.4 in Ref. \citenum{BandsLa3Te4}).

Slightly unusual temperature dependences of $\alpha$ are observed, which are likely due to the existence of multiple conduction bands within a few tenths\cite{Ce3Te4APL} of an electron volt of the conduction band edge.  Specifically, in the $x=0$ and La = 0.33 samples a small plateau exists from $\sim$50 to 90\,K.  This feature is also observed, albeit less pronounced, in La$_3$Te$_4$.  For the $x=0.25$ sample, an apparent enhancement in $\alpha$ is observed near room temperature, and this is most likely associated with the probing of the heavy conduction bands as the thermal energy increases for this composition.

\begin{figure}
	\centering
\includegraphics[width=3.in]{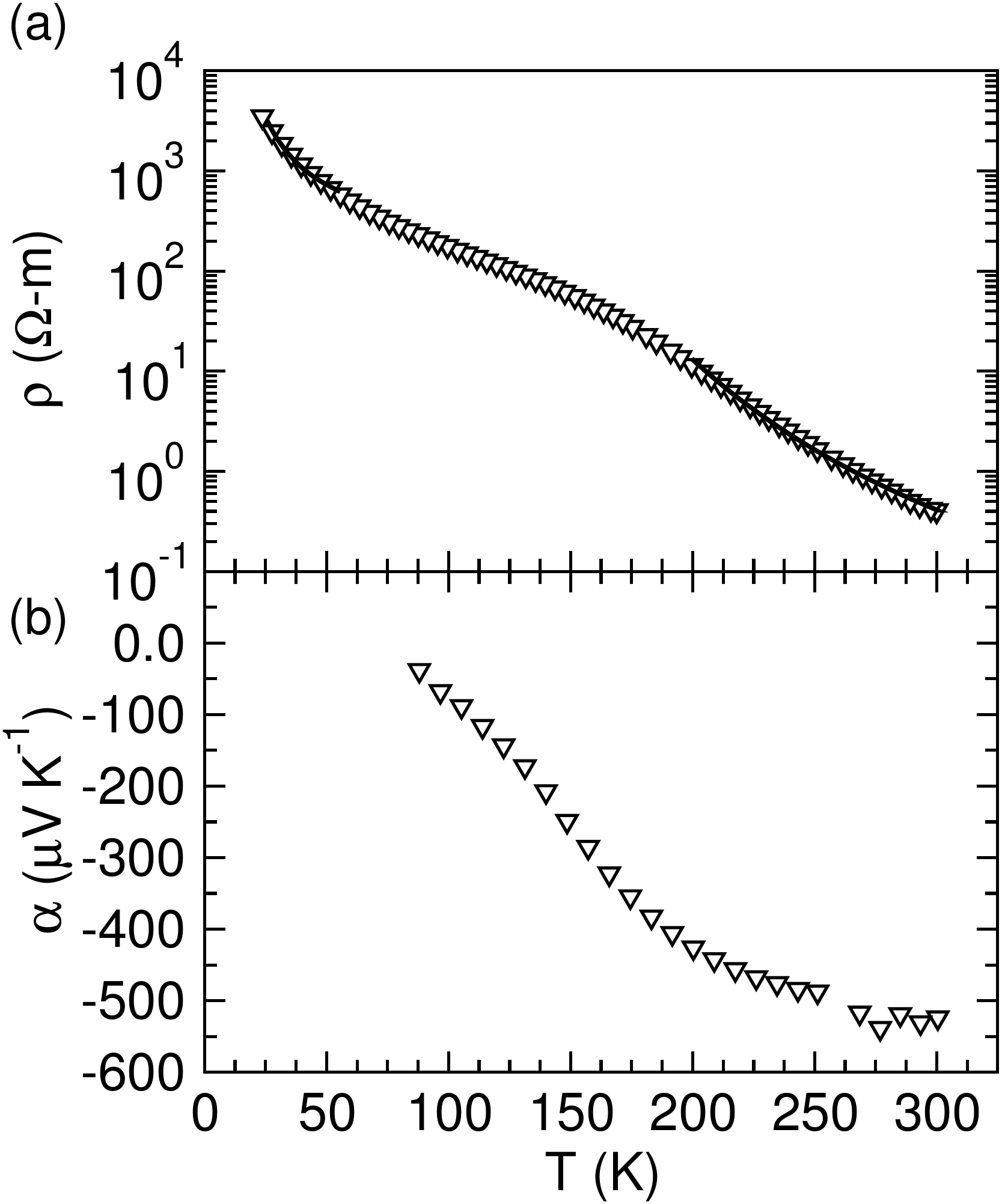}
\caption{The electrical resistivity and Seebeck coefficient of the $x=0.33$ sample are consistent with a very low extrinsic carrier concentration, with fits (solid curves) to the electrical resistivity yielding activation energies of 0.18\,eV between 200 and 300\,K, and 0.0067\,eV from 25 to 55\,K.}
	\label{fig:elec2}
\end{figure}

Above $\sim$150\,K, the temperature dependence of the electrical resistivity (Fig.\,\ref{fig:elec}a) is somewhat weaker than typically observed in metals or doped semiconductors.  In this temperature range, the Hall mobility is described by a simple power law $\mu_H$ $\propto$ $T^y$ with $-0.45 < y < -0.3$.  Similar behavior is observed for $\rho(T)$ in La$_3$Te$_4$, and thus this behavior is not likely to be associated with moments on the magnetic Ce$^{3+}$ ions. In an extrinsically doped semiconductor where acoustic phonon scattering dominates the carrier mobility, simple parabolic band theories predict $-1.5 \le\ y \le -1$, and when ionized impurities limit the mean free path $y \sim +1.5$.  One could perhaps argue that the observed behavior is a competition between these two common scattering mechanisms.  Considering the energy scale relative to the Debye temperatures, which are observed to be on the order of 170\,K, it also seems plausible that optical phonon scattering is important.  At temperatures greater than the Debye temperature, the scattering of carriers by optical phonons leads to a value of $y = -0.5$.\cite{Fistul}  There is no clear trend in the experimental $y$ with composition.  This suggests that the vacancies are not strongly influencing the carrier mobility.  The mobility is actually larger for the defect-rich $x=0.25$ sample.  This increase in $\mu_H$ may be due to the reduced effective mass at this carrier density, and the scattering of electrons by acoustic phonons also results in larger $\mu$ for lower carrier densities and lower effective mass.

The electrical resistivity and Seebeck coefficient of Ce$_{2.67}$Te$_4$ are shown in Figure \ref{fig:elec2}.  For this near-insulating composition, the electrical resistivity displays activated behavior, with three distinct regions being observed.  From 200 to 300\,K, the data can be described by $\rho=\rho_0$Exp[$E_a/kT$] with an activation energy $E_a$ of $\sim$0.18\,eV.  If this behavior is related to the activation of electron-hole pairs across the energy gap, a band gap of $E_g = 2E_a \approx$\,0.36\,eV is obtained.  This is somewhat smaller than expected, as the band gap has been calculated to be 1.06\,eV.\cite{Ce3Te4APL}  At low temperatures, the activation energy decreases dramatically and there is a large temperature range where the data are not described by simple activated behavior.  Between 25 and 50\,K, the resistivity data can be described by activated conduction with an activation energy of 6.7\,meV.  Such a small activation energy may be associated with a mobility edge, and relatively similar activation energies were obtained by Cutler and Mott when they investigated Anderson localization in Ce$_{3-x}$S$_4$.\cite{CutlerMott}

The Seebeck coefficient of Ce$_{2.67}$Te$_4$ displays behavior typical of a lightly-doped semiconductor, and does not reveal any significant influence of minority-carrier compensation up to 200\,K.  At 300\,K, though, the Seebeck coefficient may be reaching a maxima, which could then be related to the energy gap via $E_g$=2e$\alpha_{max}T_{max}$.\cite{EgEstimate}  Interestingly, using the $\alpha$ data at room temperature yields $E_g$=0.32\,eV, which is close to the energy gap obtained from $\rho(T)$ near room temperature.

It has been proposed that the Ce $f$-levels result in a large peak in the density of states $\sim$0.2\,eV above the Fermi level for Ce$_3$Te$_4$ ($x=0$), and that this electronic structure may lead to very large thermoelectric efficiency at high temperatures.\cite{Ce3Te4APL}  However, similar thermoelectric properties are observed in Ce$_3$Te$_4$ and La$_3$Te$_4$.  While the results presented here only cover 300\,K and below, it appears Zhuze et al. considered these materials at high temperatures and came to similar conclusions.\cite{Zhuze70}  Usually, band structure features that influence thermoelectric performance are observed at low temperatures because thermal disorder tends to broaden band structure features at higher temperatures.  Also, as the Fermi distribution function broadens with increasing temperature, the chemical potential in Ce$_3$Te$_4$ will move towards the conduction band edge (further from the peak in the density of states).  It seems unlikely that the high-energy $f$ states will influence the transport in Ce$_3$Te$_4$ more dramatically than in La$_3$Te$_4$, though high temperature measurements would be necessary to draw a final conclusion.  It would be interesting to see the results of Boltzmann transport calculations on a first principles electronic structure, which would provide a nice comparison between these materials if correlations are properly considered.

\begin{figure}
	\centering
\includegraphics[width=3.in]{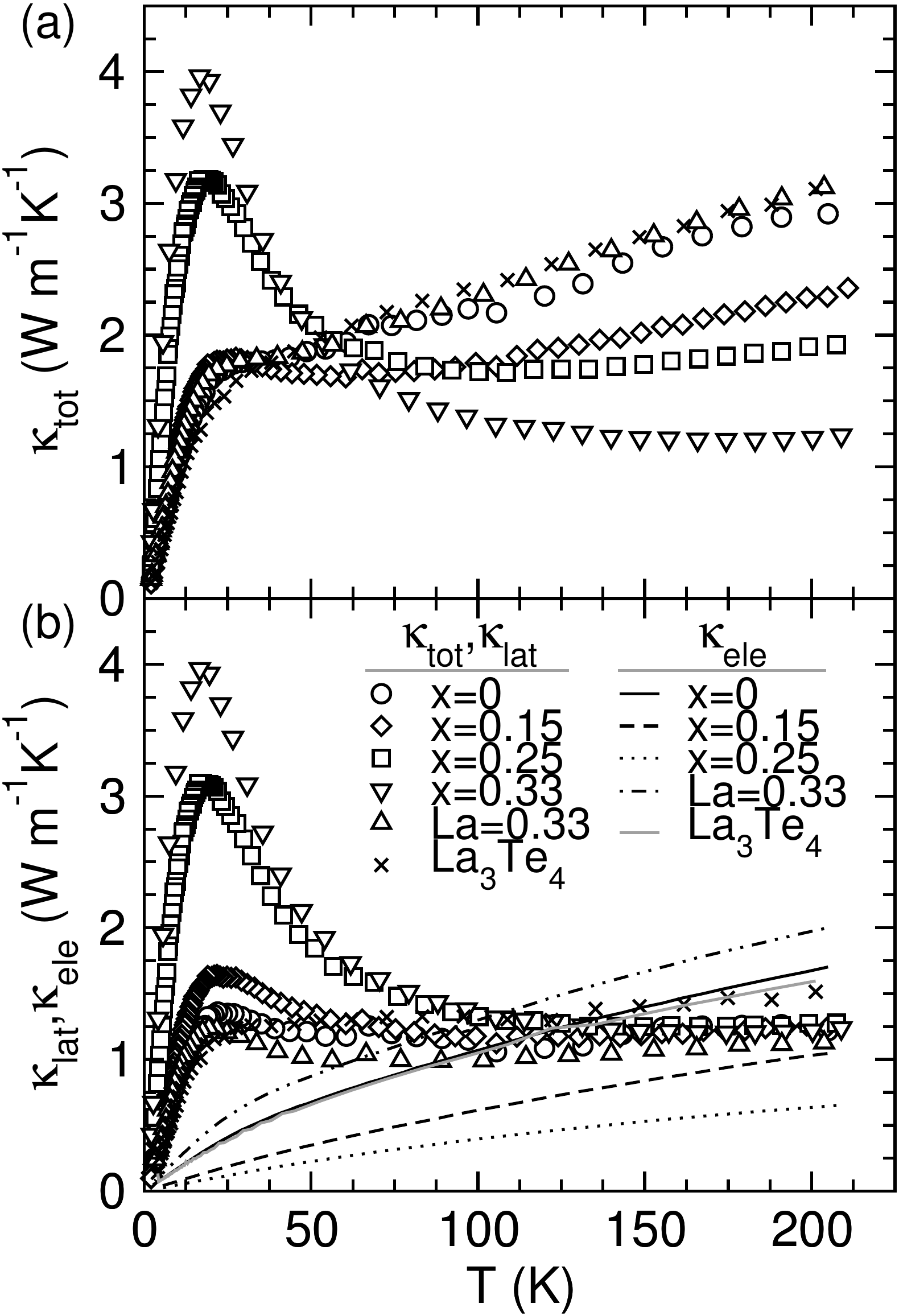}
\caption{The thermal conductivity of Ce$_{3-x}$Te$_4$ is relatively low, and the low temperature peak is found to increase with increasing cerium vacancy concentration.  This trend with $n$ reveals a strong scattering of phonons by electrons.}
	\label{fig:therm}
\end{figure}

The thermal conductivity data are presented in Figure \ref{fig:therm}, with the total measured ($\kappa_\mathrm{{tot}}$) thermal conductivity shown in Fig.\,\ref{fig:therm}a and the calculated electronic contribution $\kappa_\mathrm{{ele}}$ and lattice thermal conductivity $\kappa_\mathrm{{lat}}$ shown in Fig.\,\ref{fig:therm}b.  Similar values are obtained for Ce$_3$Te$_4$ and La$_3$Te$_4$, and thus it appears the particular rare-earth ion has relatively little influence on the thermal conductivity.  The electronic contribution is significant by 100\,K, especially for the most metallic samples, leading to an increase in $\kappa_\mathrm{{tot}}$ at high temperatures.  This is one reason that $zT$ does not optimize near the $x=0$ composition in $R_{3-x}$$Ch_4$ compounds.

Both the total thermal conductivity and lattice contribution show a clear trend with rare-earth content, particularly at low temperatures.  The low-temperature maximum is largest for the lowest rare-earth content, and smallest for the highest rare-earth content (highest $n_H$).  This is due to the scattering of phonons by electrons, which is generally large in high effective mass compounds.\cite{ZimanEP1,ZimanEP2,FeSi_Sales2011} Vacancies (point defects) are more efficient at scattering short-wavelength, optical phonons\cite{KlemensMass} and thus generally do not dominate phonon transport at low temperatures where the optical phonons are frozen-out.  As discussed above, the average grain size in the insulating $x=0.33$ sample is actually smaller than in the $x=0$ sample, and thus particle size is clearly not producing the observed trends.  Also, the small change in the Debye temperatures (Table \ref{tab:props}) is not enough to account for such a large difference in the low temperature maxima.   By high temperatures, phonon-phonon scattering dominates and the values of the lattice thermal conductivity begin to converge around 150\,K. 

The lattice contributions shown in Fig.\,\ref{fig:therm}b were obtained via $\kappa_\mathrm{{lat}}$=$\kappa_\mathrm{{tot}}$-$\kappa_\mathrm{{ele}}$, where $\kappa_\mathrm{{ele}}$ is estimated via the Wiedemann-Franz law $\kappa_\mathrm{{ele}}$=$L T/\rho$. The Lorenz numbers $L$ were calculated assuming acoustic phonon scattering limits the carrier mean free path, and a single parabolic band model was utilized.  This is certainly an approximation, especially considering the complex band structure and scattering mechanisms in these materials.  The calculated values of $L$ are all within 2\% of the degenerate limit of 2.44$\times 10^{-8}$W$\Omega$K$^{-2}$ (due to the low temperature and relatively large carrier densities).  These values are generated after obtaining the electrochemical potentials from the experimental Seebeck coefficients using the same approximations.  A more detailed discussed of the parabolic band model utilized can be found in Ref.\citenum{BGGMay}.  Note that the electronic contribution is negligible for the insulating $x=0.33$ sample.

\subsection{Magnetism}

\begin{figure}
	\centering
\includegraphics[width=3.in]{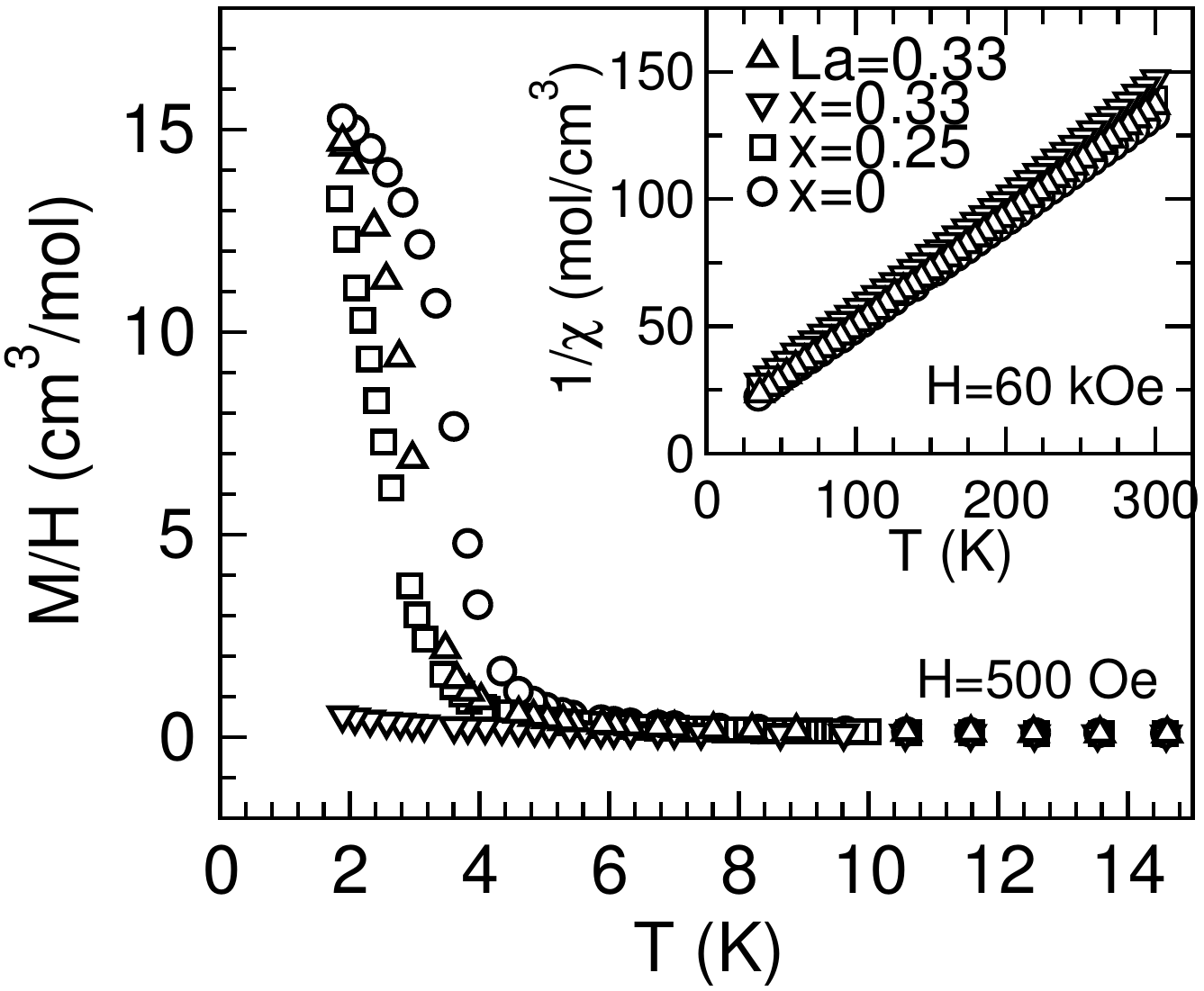}
\caption{Temperature dependence of the induced moment \textbf{M} divided by applied field of 500\,Oe suggests magnetic ordering occurs in all samples except for the $x=0.33$ sample. The inset shows the expected Curie-Weiss behavior of the magnetic susceptibility $\chi$ at higher temperatures.}
	\label{fig:chi}
\end{figure}

\begin{figure}
	\centering
\includegraphics[width=3.in]{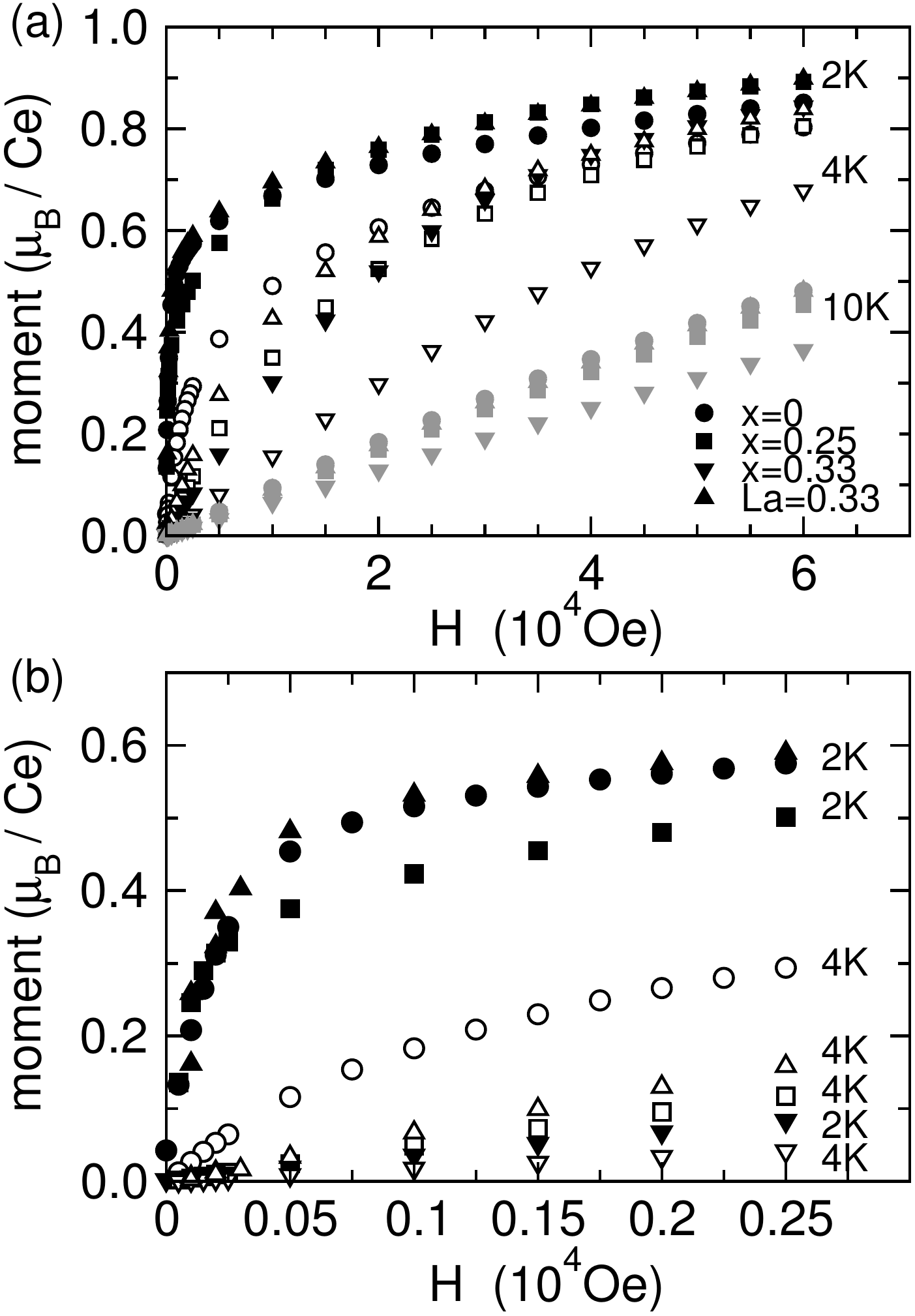}
\caption{Dependence of the induced moment on the applied field shown on two scales to reveal the (a) high field and (b) low field behavior for $T \le 10$\,K. The moment is clearly suppressed in the insulating $x=0.33$ sample.  Ce site disorder itself has little influence on the induced moment, as demonstrated by the La-substituted sample.}
	\label{fig:MH}
\end{figure}

The magnetization data for Ce$_{3-x}$Te$_4$ with $x=0,0.25,0.33$ and Ce$_{2.67}$La$_{0.33}$Te$_4$ (labeled La=0.33) are shown in Figures \ref{fig:chi} and \ref{fig:MH}.  The magnetization measurements suggest ferromagnetic ordering of Ce ions.  While Ce moments often order antiferromagnetically, ferromagnetic ordering of Ce moments is known to exist in isostructural Ce$_{3-x}$S$_4$,\cite{Ho82CeS} as well as in the cerium pnictides with the anti-Th$_3$P$_4$ structure type, such as Ce$_4$Bi$_3$.\cite{Alonso92}  

The Curie temperature $T_C$ generally decreases with decreasing cerium content, as shown in Table \ref{tab:props}.  This long-range order is suppressed to below 2\,K in the insulating $x=0.33$ sample, as highlighted by the reduced moment \textbf{M} at low temperatures.  The values of \textbf{M/H} nearly converge at low temperatures for the samples with higher Ce content (higher carrier densities).  Similarly, the moment rises sharply with field at low $T$ in the metallic samples (Fig.\,\ref{fig:MH}b), but \textbf{M(H)} increases linearly for the insulating sample.

The trends between the magnetization data and composition suggest that long-range order is driven by RKKY-type indirect exchange via the conduction electrons.    To isolate the effect of carrier concentration from the site disorder associated with cerium vacancies, a sample of nominal composition Ce$_{2.67}$La$_{0.33}$Te$_4$ was produced.  This sample has a high carrier concentration, as inferred from the low electrical resistivity, but Ce site disorder similar to insulating Ce$_{2.67}$Te$_4$.  As shown in Table \ref{tab:props}, the Curie temperature is only slightly suppressed in this La-substituted sample.  This confirms that the loss of free electrons is primarily responsible for the suppression of magnetic order in Ce$_{2.67}$Te$_4$. Small decreases in the electrical resistivity corresponding to the magnetic transitions could be identified in the $x=0,0.15$ and La=0.33 samples.

In the metallic samples ($x\le0.25$, La=0.33), the moment is nearly saturated at the highest fields, reaching $\sim$0.8-0.9\,$\mu_B$/Ce at 6\,T.  As discussed below, the heat capacity $C_P$ data for $x=0$ suggest the Ce ions are in a J=1/2 doublet state due to crystal field effects.  A saturation moment of 0.71\,$\mu_B$/Ce is expected for this doublet, and the slightly higher values observed are likely associated with the population of higher energy states due to the applied fields.

Above 30\,K, the data are well described by a modified Curie-Weiss law, as suggested by the linear behavior of $1/\chi$ versus $T$ in the inset of Fig.\,\ref{fig:chi}.  The calculated effective moments were generally near the 2.54$\mu_B$ expected for Ce$^{3+}$ (see Table \ref{tab:props}), revealing that the crystal field splitting is small relative to the thermal energy in this temperature range.  The Weiss temperatures were all negative, with values between -17.5 and -26.3\,K obtained, suggesting antiferromagnetic coupling tendencies at higher temperatures.  Fitting the susceptibility data between 10 and 30\,K yields reduced effective moments (Table \ref{tab:props}) consistent with the dominance of a doublet ground state, and Weiss temperatures were generally small and positive.  The susceptibility data utilized to obtain these results were collected upon cooling in an applied field of 6\,T.

\subsection{Specific Heat}

\begin{figure}
	\centering
\includegraphics[width=3.in]{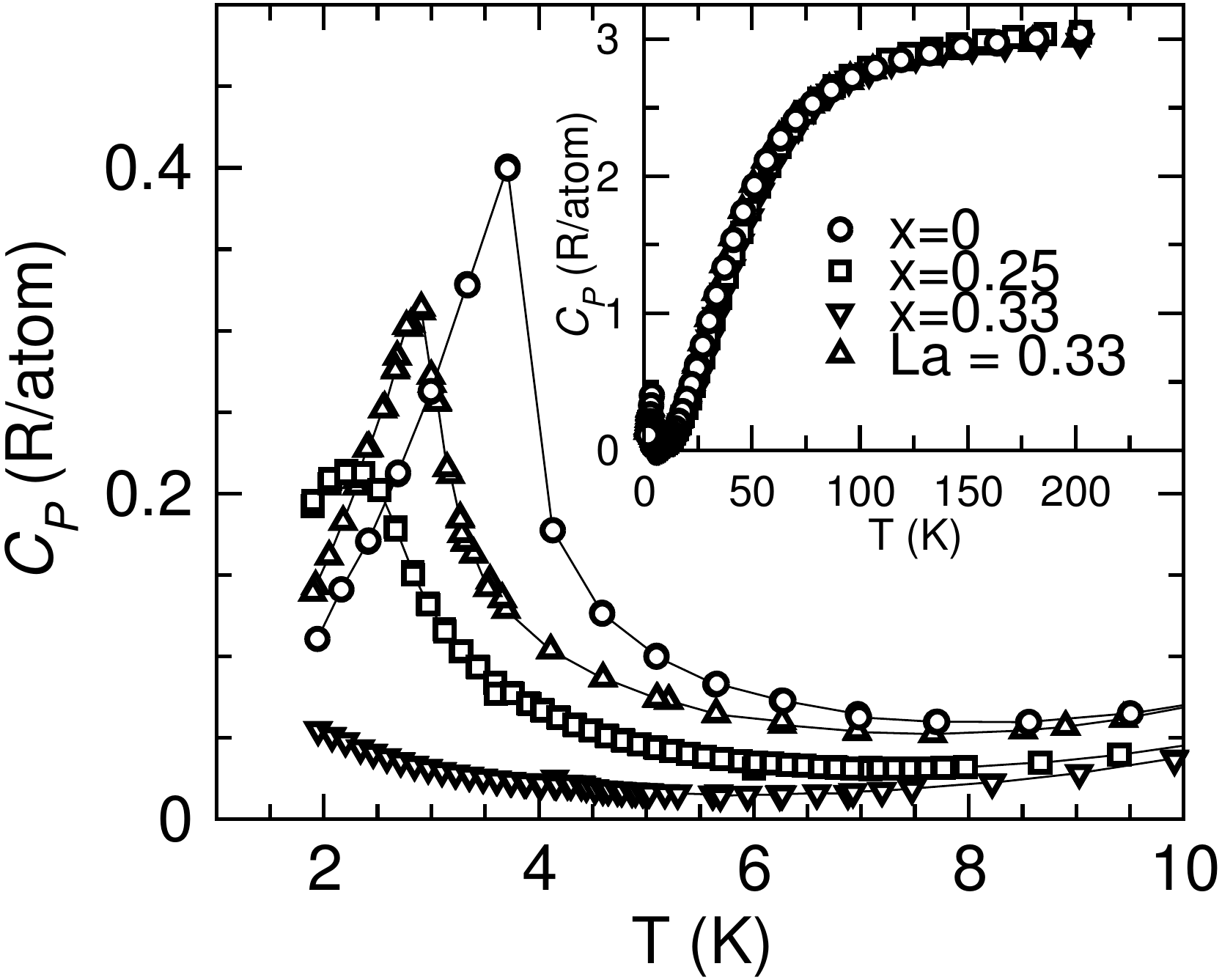}
\caption{The specific heat of Ce$_3$Te$_4$ reveals a strong anomaly associated with the ferromagnetic ordering of Ce moments near 4\,K, and the anomaly is suppressed to below 2\,K for Ce$_{2.67}$Te$_4$.}
	\label{fig:Cp}
\end{figure}

The specific heat capacity $C_P$ of Ce$_{3-x}$Te$_4$ samples is shown in Figure \ref{fig:Cp}.  These measurements, which are performed in zero magnetic field, are highly sensitive to the ordering of Ce moments.  As such, the suppression of the magnetic order with decreasing Ce content is clearly observed.  For the fully-filled Ce$_3$Te$_4$ compound, a large peak in $C_P$ is observed near 3.7\,K.  Only a weak up-turn in $C_P$ is observed in the $x=0.33$ sample, suggesting the magnetic order is suppressed to very low temperatures in this sample.  This suppression of magnetic ordering is primarily associated with the change of carrier concentration and not the cerium-site disorder.  The relatively strong peak in the disordered, but electron-rich Ce$_{2.67}$La$_{0.33}$Te$_4$ confirms that indirect exchange is dominant in this system.

The entropy change $\Delta S$ associated with the magnetic transition in Ce$_3$Te$_4$ was obtained by $\Delta S = \int_{1.95}^{15} \frac{C_{P}}{T}\mathrm{d}T$.  The baseline for this integration was the heat capacity of La$_3$Te$_4$, and a 12\,T magnetic field was applied to suppress the superconducting transition.\cite{OlivierLaTe}  For the $x=0$ sample, $\Delta S$ was calculated to be $\sim$4.7J/mol-Ce/K = 0.82RLn2 (on a per mol cerium basis).  This suggests the crystal field splitting leads to the J=1/2 doublet state for Ce$^{3+}$, for which the expected entropy change would be RLn2.  Similar behavior has been observed in Ce$_{3-x}$S$_4$.\cite{Ho82CeS}

Debye temperatures $\Theta_D$ were obtained by fitting the specific heat capacity between 20 and 200\,K to the simple Debye model, 

\begin{equation}
C = 9 R N_{at} \left(\frac{T}{\Theta_D}\right)^3 \int^{\Theta_D/T}{\frac{y^4 e^y}{(e^y-1)^2}\mathrm{d}y}.
\label{eqn:Cp}
\end{equation}

\noindent Nominal compositions were utilized to estimate the number of atoms per formula unit $N_{at}$. The obtained values of $\Theta_D$ (Table \ref{tab:props}) trend with the cerium content, with lower Debye temperatures obtained for higher cerium content.  Similar trends were also observed in La$_{3-x}$S$_4$ and La$_{3-x}$Te$_4$,\cite{Ikeda82Debye,Smirnov75,OlivierLaTe} and this softening of the phonons is caused by the increased density of states at the Fermi energy as rare-earth content (or $n$) increases.\cite{OlivierLaTe}   The Debye model does a fair job of describing the data below 50 or 80\,K, and generally underestimates the data at higher temperatures for Ce$_{3-x}$Te$_4$.  The insulating $x=0.33$ sample is better described at higher temperatures.  Therefore, the lack of an electronic component is likely the main reason for the discrepancy between Eqn. \ref{eqn:Cp} and the data at higher temperatures.  The dilation term that accounts for differences between the constant pressure measurement and the constant volume calculation is likely important at higher temperatures than those considered here.  The magnetic transitions prevent the low-$T$ Debye temperature and Sommerfeld coefficients from being obtained.

\section{Summary}

The compounds Ce$_{3-x}$Te$_4$ are found to order ferromagnetically at temperatures below $\sim$4\,K, with the ordering temperature generally decreasing with decreasing carrier density and/or cerium content.  The magnetic ordering is suppressed to below 2\,K in the vacancy rich, insulating Ce$_{2.67}$Te$_4$ composition.  This trend, and the observation of magnetic ordering near 3\,K in Ce$_{2.67}$La$_{0.33}$Te$_4$ reveal that RKKY indirect exchange drives the ordering of magnetic moments.  Heat capacity and magnetic susceptibility measurements are consistent with a J=1/2 doublet at low temperatures due to crystal field splitting.  The thermoelectric transport properties in Ce$_3$Te$_4$ are found to be similar to those of La$_3$Te$_4$ below room temperature.  It seems unlikely that Ce$_{3-x}$Te$_4$ compounds will have significantly greater thermoelectric performance than La$_{3-x}$Te$_4$ compounds at high temperatures.

\section{Acknowledgements}
 
This work was supported by the U. S. Department of Energy, Office of Basic Energy Sciences, Materials Sciences and Engineering Division. This research was also partially supported by ORNL SHaRE, sponsored by the Division of Scientific User Facilities, Office of Basic Energy Sciences, U.S. Department of Energy.

\end{document}